\documentclass
{amsart}
\usepackage{amssymb,amsmath,graphicx}
\theoremstyle{definition}

\theoremstyle{remark}

\numberwithin{equation}{section}

\begin{document}
\title [Thresholds of Asymptotic Freedom] {SUGRA Interactions at Thresholds of Asymptotic Freedom}
\author{J. Towe}
\address{Department of Physics, The Antelope Valley College, Lancaster, CA 93536}%
\email{jtowe@avc.edu}\
\begin{abstract}
A generic, heterotic string theory is uniquely reduced to the
standard model in terms of a geometry that transcends
compactification. This device also extends the standard model to
embrace three generations of fermions, including a left-handed
strange quark that is devoid of strangeness and three right-handed
neutrinos. Finally, the proposed hypothesis indicates
supergravitationally mediated quark-lepton transitions that preserve
baryon structure, departing from the SUSY GUT tradition that
predicts proton decay.


\end{abstract} \maketitle

\section {A Unique Reduction to the Standard Model}\label{S:intro}
The current status of high energy physics confronts two significant
problems: that SUSY GUT theories predict a proton decay, which is
not observed, and that heterotic string models (the only finite,
physical models) of supergravity do not uniquely reduce to the
standard model at SM scale. Both problems are addressed in terms of
a geometry, which transcends compactification by preserving the
geodesic nature of world tube coupling.
\par
A pure supergravitational transition at string scale is envisioned
as an absorption-radiation event that is encountered by a closed
string of spin 3/2, as that string sweeps out a geodesic world tube
in 10-spacetime. Each absorption-radiation event involves the
absorption of a string of spin 2 and the simultaneous radiation of a
second, generally distinct string of spin 2 through the mediation of
a graviton vertex operator (GVO). The constraint of simultaneity is
necessary to conserve intrinsic angular momentum along the world
tube of spin 3/2. Two classes of spin-2 fields and two classes of
GVOs are postulated. A Class I GVO is identified with the absorption
of a Class I string of spin 2 and the radiation of a Class II string
of spin 2. A Class II GVO is identified with the opposite order of
absorption and radiation.
\par
It is observed that GVOs experience 'geodesic coupling' at string
scale, where geodesic coupling refers to coupling in terms of a
world tube that is swept out by a string of spin 3/2 (closed string
analogue of a gauge field of local supersymmetry). In this context a
transcendental geometry is defined as consisting of world tubes of
spin 3/2 that map under compactification onto spin 3/2 geodesic
couplings of GVO particle analogues (also to be known as GVOs) and
of their spin 2 superpartners. Because compactification is defined
in terms of an $E_{8}$ symmetry, which breaks to yield SU(5)XSU(3)
(in both observable and hidden sectors), a spin 3/2 geodesic
coupling is defined as one that preserves SU(5)XSU(3) as well as
local supersymmetry and the spin-(3/2) nature of the gauge coupling.
If SU(5) is defined so that the conserved parameters are $I_{3}$, Y,
fermion number and generation, then the proposed hypothesis yields
three fundamental representations of SU(5) in terms of GVOs, which
correspond to three orientations of the $I_{3}$ axis about the Y
axis, forming a representation of SU(3) on a plane that is
orthogonal to the orientations of the SU(5) symmetries. Thus if one
can constrain the spin 1 sector of each GVO to consist exclusively
of a photon, then up to constant scale factors,
three generations of quarks and three generations of leptons are
established. This reduction of heterotic string theory to the
standard model should be compared with a reduction that was obtained
from more traditional considerations [W. Buchm$\ddot{u}$ller,2005].
\par
The coordinates and corresponding gauge particles of one orientation
of the $I_{3}$XY grid about the Y axis are:
\begin{equation} \label{E:int}
I_{3}=1/2, Y=1/3: U_{L}, \nu^{e^{-}}_{L}
\end{equation}
\begin{equation} \label{E:int}
I_{3}=-1/2, Y=1/3: D_{L}, e^{-}_{L}
\end{equation}

\begin{equation} \label{E:int}
I_{3}=0, Y=2/3: \overline{S}_{L}, \mu^{+}_{L}
\end{equation}

\begin{equation} \label{E:int}
I_{3}=1/2, Y=-2/3: U_{R}, \nu^{e^{-}}_{R}
\end{equation}
\begin{equation} \label{E:int}
I_{3}=-1/2, Y=-2/3: D_{R}, e^{-}_{R}.
\end{equation}
The coordinates and corresponding gauge particles of a second
orientation, $2\pi/3$ from the initial orientation, are:
\begin{equation} \label{E:int}
I_{3}=1/2, Y=1/3: C_{L}, \nu^{\mu^{-}}_{L}
\end{equation}
\begin{equation} \label{E:int}
I_{3}=-1/2, Y=1/3: S_{L}, \mu^{-}_{L}
\end{equation}

\begin{equation} \label{E:int}
I_{3}=0, Y=2/3: \overline{S}_{L}, \mu^{+}_{L}
\end{equation}

\begin{equation} \label{E:int}
I_{3}=1/2, Y=-2/3: C_{R}, \nu^{\mu^{-}}_{R}
\end{equation}
\begin{equation} \label{E:int}
I_{3}=-1/2, Y=-2/3: S_{R}, \mu^{-}_{R}.
\end{equation}
Finally, the coordinates and gauge particles of a third orientation,
$4\pi/3$ from the initial orientation, are:
\begin{equation} \label{E:int}
I_{3}=1/2, Y=1/3: T_{L}, \nu^{\tau^{-}}_{L}
\end{equation}
\begin{equation} \label{E:int}
I_{3}=-1/2, Y=1/3: B_{L}, \tau^{-}_{L}
\end{equation}

\begin{equation} \label{E:int}
I_{3}=0, Y=2/3: \overline{S}_{L}, \mu^{+}_{L}
\end{equation}

\begin{equation} \label{E:int}
I_{3}=1/2, Y=-2/3: T_{R}, \nu^{\tau^{-}}_{R}
\end{equation}
\begin{equation} \label{E:int}
I_{3}=-1/2, Y=-2/3: B_{R}, \tau^{-}_{R}.
\end{equation}

In all cases, the Y coordinate of the quark and lepton are
calculated in terms of the equation
\begin{equation} \label{E:int}
Y=B+S+QHC+LHC,
\end{equation}
and the $I_{3}$ coordinate of the quark and lepton are calculated in
terms of the equation:
\begin{equation} \label{E:int}
I_{3}=Q+(1/2)QHC+(1/2)LHC+(1/2)LIC-(1/2)Y,
\end{equation}
where QHC refers to 'quark hypercharge constant', LHC refers to
'lepton hypercharge constant', LIC refers to 'lepton isospin
constant' and LIC+LHC refers to the sum. The values of these
constant scale factors that are compatible with both symmetry and
internal consistency are:
\begin{equation} \label{E:int}
QHC(RH non-strange quark)=-1
\end{equation}
\begin{equation} \label{E:int}
QHC(everything else)=0
\end{equation}
\begin{equation} \label{E:int}
LHC(LH lepton)=1/3,
\end{equation}
\begin{equation} \label{E:int}
LHC(RH lepton)=-2/3,
\end{equation}
\begin{equation} \label{E:int}
(LIC+LHC)(LH lepton)=4/3
\end{equation}
\begin{equation} \label{E:int}
(LIC+LHC)(LH anti-lepton)=-4/3
\end{equation}
\begin{equation} \label{E:int}
(LIC+LHC)(RH lepton)=1/3
\end{equation}
and
\begin{equation} \label{E:int}
(LIC+LHC)(RH anti-lepton)=-1/3.
\end{equation}
This result should be compared with the traditional standard model
and with the extension of this model that includes three fermionic
generations [D. Nordstrom, 1992].


\par
The issue of proton decay is now addressed. It is generally accepted
that quarks coalesce to form triplets at about 1 TeV, and that
supersymmetry is intact at scales just above 1 TeV. It is therefore
argued that the supergravitational activity, which has been
discussed, is indeed realized at scales below compactification
scale, and specifically at scales around 1 TeV, by interactions
between fields of spin 2 and valance quarks that reside within
baryons of spin 3/2. These are quarks that are experiencing
asymptotic freedom. It can be shown that such interactions include
quark-lepton interactions that preserve baryon structure; but first
the nature of the spin-2 fields that emerge from the proposed theory
must be considered. The next section will describe such fields.

\section{Specific Spin-2 Fields}\label{S:intro}
The formulation of an appropriate spin-2 field involves a GVO
\begin{equation} \label{E:int}
OP_{e^{-}_{L}}\equiv\gamma_{L}\otimes\overline{e^{-}}_{R}
\end{equation}
and a corresponding spin-(3/2) field
\begin{equation} \label{E:int}
\psi_{D_{L}}=\overline{U}_{L}^{\prime}D_{L}U_{L}
\end {equation}
where a bar indicates an anti-field and a prime indicates
anti-alignment of an anti-field.
If the entities 2.1 and 2.2 coalesce, then the result is clearly the
spin-2 field
\begin{equation} \label{E:int}
OP_{e^{-}_{L}}\otimes\psi_{D_{L}}\equiv G_{L}^{A}
\end {equation}
or
\begin{equation} \label{E:int}
\gamma_{L}\otimes\overline{e^{-}}_{R}\otimes
\overline{U}_{L}^{\prime}D_{L}U_{L} \equiv G_{L}^{A}
\end {equation}
which describes a spin-2 resultant field that retains the color of
$D_{L}$, retains the charge $Q(D_{L})+Q(\overline{e^{-}_{L}})=2/3$
and the $I_{3}$ number: $I_{3}$=0.
\par
A second spin-2 field class is produced when the GVO
\begin{equation} \label{E:int}
OP_{{D}_{L}}\equiv\gamma_{L}\otimes\overline{D}_{R}
\end {equation}
interacts with the same spin-(3/2) field as that described by
Equation 2.2
($\psi_{D_{L}}=\overline{U}_{L}^{\prime}D_{L}U_{L}$).
In this case one obtains
\begin{equation} \label{E:int}
OP_{D_{L}}\otimes\psi_{D_{L}} \equiv g_{L}^{A}
\end {equation}
or
\begin{equation} \label{E:int}
\gamma_{L}\otimes\overline{D}_{R}\otimes
\overline{U}_{L}^{\prime}D_{L}U_{L} \equiv g_{L}^{A}.
\end {equation}
The latter spin-2 field is clearly characterized by zero mass, zero
charge, is colorless and is also characterized by $I_{3}$=0. The
spin-2 fields $G_{L}$ and $g_{L}$ will subsequently be known as
Class I, Type A and Class II, Type A.
\par
Let us now consider the two GVOs
\begin{equation} \label{E:int}
OP_{\nu_{e^{-}}}\equiv\gamma_{L}\otimes\overline{\nu^{e^{-}}}_{R}
\end {equation}
and
\begin{equation} \label{E:int}
OP_{U_{L}}\equiv\gamma_{L}\otimes\overline{U}_{R}
\end {equation}
which interact with the spin-(3/2) field
\begin{equation} \label{E:int}
\psi_{U_{L}}\equiv\overline{D}^{\prime}_{L}U_{L}D_{L}
\end {equation}
to respectively produce the spin-2 fields
\begin{equation} \label{E:int}
\gamma_{L}\otimes\overline{\nu^{e^{-}}}_{R}\otimes
\overline{D}_{L}^{\prime}U_{L}D_{L} \equiv G^{B}_{L}
\end {equation}
and
\begin{equation} \label{E:int}
\gamma_{L}\otimes\overline{U}_{R}\otimes
\overline{D}_{L}^{\prime}U_{L}D_{L} \equiv g^{B}_{L}
\end {equation}
(Note that the quantum numbers, other than mass (which is presumably
absent in contexts where SUSY is unbroken), of $G^{B}_{R}$ and
$g^{B}_{R}$ are the same as those that characterize $G^{A}_{R}$ and
$g^{A}_{R}$.)
\par
Thirdly, let us consider the two GVOs
\begin{equation} \label{E:int}
OP_{e^{-}_{R}}\equiv\gamma_{R}\otimes\overline{e^{-}}_{L}
\end {equation}
and
\begin{equation} \label{E:int}
OP_{D_{R}}\equiv\gamma_{R}\otimes\overline{D}_{L}
\end {equation}
that interact with the spin-(3/2) field
\begin{equation} \label{E:int}
\psi_{D_{R}}=\overline{D}^{\prime}_{R}D_{R}D_{R}
\end {equation}
to respectively produce the spin-2 fields
\begin{equation} \label{E:int}
\gamma_{R}\otimes\overline{e^{-}}_{L}\otimes
\overline{D}_{R}^{\prime}D_{R}D_{R} \equiv G^{C}_{R}
\end {equation}
and
\begin{equation} \label{E:int}
\gamma_{R}\otimes\overline{D}_{L}\otimes
\overline{D}^{\prime}_{R}D_{R}D_{R} \equiv g^{C}_{R}
\end {equation}
(Again, the quantum numbers of $G^{C}_{R}$ and $g^{C}_{R}$ are the
same as those that characterize $G^{A}_{R}$ and $g^{A}_{R}$. As
indicated above, analogues of $G^{A}, G^{B}, G^{C}, g^{A}, g^{B}$
and  $g^{C}$ exist for all three generations [J. Towe, 2003].)
\par
Now that spin-2 fields have been assigned specific structure, one
can consider a realization of the theory of pure supergravity which
impacts the standard model. Again the proposed realization occurs at
approximately 1 TeV, where quarks are bound into triplets. Because
supersymmetry remains intact at just above 1 TeV, it is argued that
the proposed supergravitational interactions can occur between
fields of spin 2 and baryons of spin 3/2. Specifically, such
interactions are regarded as occurring between the proposed fields
of spin-2 and valance quarks that reside within baryons of spin 3/2.
These are quarks that are experiencing asymptotic freedom. Such
interactions can also occur within baryons of spin 1/2 provided that
mediations are complemented by theoretically provided photons (this
to preserve local supersymmetry). Before considering the proposed
interactions let us summarize the above described nature of the
spin-2 fields and the supergravitational couplings that were
prescribed by this section. Spin 2 fields occur in two classes: a
class that consists of elements of charge 2/3 that are characterized
by a single color and an $I_{3}$ value of zero; and a class that
consists of elements that are charge-less, colorless and also
characterized by $I_{3}$=0. There are three types of
supergravitational couplings per generation: one for each pair of
GVOs that share a common value of $I_{3}$ and Y; i.e. one for each
type of SUGRA interaction. Interaction types are designated A, B and
C. Prototypes of the interactions that are proposed for the 1 TeV
energy scale will now be considered.
\section {Supergravitationally Mediated Quark-lepton Transitions}\label{S:intro}
An example of the Type A interaction occurs when a left-handed down
quark absorbs a spin-2 anti-field of class I, Type A, also
designated $\overline{G}_{R}^{A}$ and radiates a spin-2 anti-field
of class II, Type A, also designated $\overline{g}_{R}^{A}$,
producing a left-handed electron. This transition is quickly
reversed so that baryon structure is preserved.


\par
An example of a Type B interaction occurs when a left-handed up
quark absorbs a spin 2 anti-field of Class I, Type B, also
designated $\overline{G}_{R}^{B}$ and radiates a spin 2 anti-field
of Class II, Type B, also designated $\overline{g}_{R}^{B}$ to
produce a colorless, left-handed particle of spin 1/2 with a charge
of zero; i.e. a left-handed electron's neutrino.
In both cases the baryon could be a $\Delta^{0}$. An example of a
Type C interaction occurs when a right-handed strange quark absorbs
a spin 2 anti-field of Class I, Type C (also designated
$\overline{G}_{R}^{C}$), generation II, and a radiates a spin 2
anti-field of Class II, Type C (also designated
$\overline{g}_{R}^{C}$), generation II to produce a right-handed
muon:
Here the baryon could be an $\Omega$ (triplet SSS). Clearly, the
proposed interactions are occurring to all generations of quarks,
and are (from symmetry considerations) generation specific.

\par
The proposed interactions can also occur within baryons of
spin-(1/2) provided that a photon is absorbed and radiated together
with each spin-2 anti-field to preserve local supersymmetry.
\par
\section{Conclusion}\label{S:intro}
$ $\\[-06pt]
It was recalled that high energy physics confronts two important
problems: that SUSY GUT theories predict a proton decay that is not
observed and that heterotic string models of supergravity do not
uniquely reduce to the standard model at SM scale. Both problems
were addressed in terms of a proposed differential geometry that
transcends compactification. Specifically, it was proposed that
geodesic world tubes be regarded as elements of a transcendent
geometry if and only if they map under compactification onto
'geodesic couplings' of GVO particle analogues, which were
subsequently referred to as GVOs. In a context where
compactification was interpreted as a breaking of $E_{8}$ which
forms SU(5)XSU(3), a geodesic coupling was defined as a coupling
that preserves SU(5)XSU(3) as well as local supersymmetry and the
spin-(3/2) nature of the gauge connection.
\par
SU(5) was defined so that the conserved parameters are $I_{3}$, Y,
fermion number and generation. In this context the proposed
hypothesis resulted in three fundamental representations of SU(5) on
$I_{3}XY$ grids that were in terms of GVOs and symmetrically
oriented about one Y axis, thereby projecting onto an orthogonal
plane as a representation of SU(3). By constraining the spin 1
sector of the GVO to consist exclusively of a photon, three
generations of quarks and three generations of leptons were
established up to constant scale factors. It was concluded that the
postulated transcendent geometry had reduced a generic, heterotic
string theory to the standard model, and had extended the standard
model to embrace three generations of fermions, including a new
quark and three right-handed neutrinos. The new quark was
characterized as a left-handed strange quark, which in the context
of what was postulated, is characterized by a baryon number of 1/3
and a hypercharge of 1/3; i.e. by a strangeness number of zero.
\par
Finally, the issue of proton decay was addressed. It was noted that
quarks come together to form triplets at about 1 TeV, and that
supersymmetry is intact at just above 1 TeV. It was therefore argued
that the supergravitational activity, which had been discussed, is
indeed realized at scales below compactification scale, and
specifically at a scale around 1 TeV, by interactions between fields
of spin 2 and valance quarks that reside within baryons of spin 3/2.
It was emphasized that these quarks are experiencing asymptotic
freedom. Before it could be demonstrated that such interactions
produce quark-lepton transitions that preserve baryon structure, it
was necessary to consider the nature of the spin-2 fields that
emerge from the proposed theory.
\par
It was observed that such fields occur in two categories: one
consisting of elements of charge 2/3, a single color and an $I_{3}$
number of zero; and a second consisting of elements that are
charge-less, colorless and also characterized by an $I_{3}$ number
of zero. In this context the three indicated types of
supergravitational interactions (per generation) were considered.
These interactions were described in some detail. It was
demonstrated that baryon structure is not disturbed and it was
observed that the same types of interactions can occur within
baryons of spin 1/2, provided that photons are absorbed and radiated
together with each spin-2 anti-field--this to preserve local
supersymmetry. In closing, the preservation of baryon structure by
interactions with the postulated gravitational radiation field was
compared with preservations of atomic structure by similar
interactions with the electromagnetic radiation field of QED.

\end{document}